\documentclass{PoS}

\title{Monopoles, topology of the Standard Model, and unification of interactions at Tev scale }

\ShortTitle{Monopoles, topology of the Standard Model, and unification of
interactions at Tev scale}

\author{\speaker{Mikhail Zubkov}%
         \thanks{ITEP, B.Cheremushkinskaya 25, Moscow, 117259, Russia}\\
        \\
        E-mail: \email{zubkov@itep.ru}}

\abstract{It is shown that unification of strong and Electroweak interactions
at Tev scale may lead to appearance of topologically stable monopoles with
masses of the order of $40$ Tev.  Those monopoles may play an
 important role in the early Universe, at finite temperature.
 They may even be condensed at high enough temperature. The lattice model for investigation
 of this phenomenon is presented.}

\FullConference{The XXV International Symposium on Lattice Field Theory\\
July 30 - August 4 2007\\
Regensburg, Germany}

\begin{document}

\section{Introduction}
Due to the so-called Hierarchy problem \cite{TEV} at the energy scale of about
$1$ Tev new physics is expected to appear. Naturalness of the theory \cite{TEV}
leads to the requirement that the ultraviolet cutoff $\Lambda$ is of the order
of $1$ TeV. It is worth mentioning that this problem appears also in lattice
nonperturbative study (see, for example, \cite{BVZ2007}). Thus at the Tev scale
some other theory should appear, which incorporates Standard Model as a low
energy approximation.

There are several patterns of unification of interactions, which were
considered up to now. Among them there are at least three examples, in which
gauge group of the Standard Model is extended already at the Tev scale. Namely,
in the so - called Little Higgs models \cite{little_higgs,TopLittleHiggs}
$SU(2)\times U(1)$ subgroup is embedded into a larger group, which is gauged
partially. The correspondent symmetry is broken at a few Tev.

The second example is the Extended Technicolor Theory.  The idea is to
construct the new Tev scale theory basing on the analogy with QCD
\cite{technicolor, WS,FS,ExtendedTechnicolor}. The new Nonabelian gauge
interaction is added with the scale $\Lambda_{TC} \sim 1$ Tev, where
$\Lambda_{TC}$ is the analogue of $\Lambda_{QCD}$. This new interaction is
called technicolor. The correspondent new fermions are called technifermions.
Breaking of the chiral symmetry in technicolor theory causes Electroweak
symmetry breaking. In order to make Standard Model fermions massive extra gauge
interaction is added, which is called Extended technicolor (ETC)
\cite{technicolor,ExtendedTechnicolor}. In this new gauge theory the Standard
Model fermions and technifermions enter the same representation of Extended
technicolor group. There is a great number of ETC models. In particular, there
exist models such that the ETC gauge group unifies strong and Electroweak
interactions at Tev scale.

The third example is the so-called Petite Unification (see, for example,
\cite{PUT,PUT1} and references therein). In the correspondent models the gauge
symmetry of the Standard Model is extended to a larger one at the Tev scale.
The resulting models have two different coupling constants correspondent to
strong and Electroweak interactions unlike Grand Unified models, in which there
is only one coupling constant and the unification is achieved at the GUT scale
$10^{15}$ Gev.

Long time ago it was recognized that the spontaneous breakdown of $SU(5)$
symmetry in the Unified model actually leads to the gauge group $SU(3)\times
SU(2) \times U(1)/Z_6$ instead of the conventional $SU(3)\times SU(2) \times
U(1)$ (see, for example, \cite{Z6} and references therein). The appearance of
the additional $Z_6$ symmetry in the fermion and Higgs sectors of the Standard
Model itself was recovered later within the lattice field theory
\cite{BVZ2003,BVZ2004, BVZ2005, BVZ2006}. Independently $Z_6$ symmetry in the
Higgs sector of the Standard Model was considered in \cite{Z6f}. The emergence
of $Z_6$ symmetry in technicolor models was considered in \cite{Z2007_2}.

Due to the $Z_6$ symmetry the gauge group of the Standard Model is either
$SU(3)\times SU(2) \times U(1)$ or $SU(3)\times SU(2) \times U(1)/{\cal Z}$,
where ${\cal Z}$ is equal to $Z_6$, or to one of its subgroups: $Z_3$ or $Z_2$.
It would be important to know is there any difference between the correspondent
models or not.

On the level of perturbation expansion those versions of the Standard Model are
identical. Here we take into account that the Standard Model describes nature
only up to the energies of about a few  Tev. This can have an effect on the
topology of the Standard Model. Namely, there may appear small regions of sizes
of the order of $1 \, {\rm Tev}^{-1}$, where the conventional fields of the
Standard Model are not defined. These regions may represent monopoles of the
unified theory. Then different versions of the Standard Model lead to different
monopole contents of the models that describe Tev - scale physics.

At zero temperature unified gauge group is broken to the Standard Model gauge
group. However, at the temperatures higher than the critical temperature  $T_c
\sim 1$  Tev this symmetry should be restored. Basing on the analogy with Nambu
monopoles \cite{BVZ2006} we expect that the monopoles of the models, which
describe Tev - scale physics, may be condensed at $T > T_c$. In order to
investigate this phenomenon we suggest to consider the lattice toy model based
on one of the Petite Unification models.

\section{Monopoles of the unified theory}
Let us now fix the closed surface $\Sigma$ in $4$-dimensional space $R^4$. For
any closed loop $\cal C$, which winds around this surface, we may calculate the
Wilson loops $\Gamma = {\rm P} \,  {\rm exp} (i\int_{\cal C} C^{\mu}
dx^{\mu})$, $U = {\rm P} \, {\rm exp} (i\int_{\cal C} A^{\mu} dx^{\mu})$, and
$e^{i\theta} =  {\rm exp} (i\int_{\cal C} B^{\mu} dx^{\mu})$, where $C$, $A$,
and $B$ are correspondingly $SU(3)$, $SU(2)$ and $U(1)$ gauge fields of the
Standard Model. In the usual realization of the Standard Model with the gauge
group $SU(3)\times SU(2) \times U(1)$ such Wilson loops should tend to unity,
when the length of $\cal C$ tends to zero ($|{\cal C}| \rightarrow 0$).
However, in the $SU(3)\times SU(2) \times U(1)/{\cal Z}$ gauge theory the
following values of the Wilson loops are allowed at $|{\cal C}| \rightarrow 0$:
\begin{eqnarray}
\Gamma &=& {\rm P} \, {\rm exp} (i\int_{\cal C} C^{\mu} dx^{\mu}) = e^{N
\frac{2\pi
i}{3}}\nonumber\\
U &=& {\rm P} \, {\rm exp} (i\int_{\cal C} A^{\mu} dx^{\mu}) =
e^{-N \pi i}\nonumber\\
e^{i\theta} &=& {\rm exp} (i\int_{\cal C} B^{\mu} dx^{\mu}) = e^{N \pi
i},\label{Sing}
\end{eqnarray}
where $N = 0,1,2,3,4,5$ for ${\cal Z}=Z_6$, $N = 0,2,4$ for ${\cal Z}=Z_3$, and
$N = 0,3$ for ${\cal Z}=Z_2$. Then the surface $\Sigma$ may carry $Z_2$ flux
$\pi [N\, {\rm mod}\,2]$ for ${\cal Z} = Z_2, Z_6$. It also may carry $Z_3$
flux $\frac{2\pi [N\, {\rm mod}\,3]}{3}$ for ${\cal Z} = Z_3, Z_6$.

Any configuration with the singularity of the type (\ref{Sing}) could be
eliminated via a singular gauge transformation. Therefore the appearance of
such configurations in the theory with the gauge group $SU(3)\times SU(2)
\times U(1)/{\cal Z}$ does not influence the dynamics.

Now let us consider an open surface $\Sigma$. Let the small vicinity of its
boundary $U(\partial \Sigma)$ represent a point - like soliton state of the
unified theory. This means that the fields of the Standard Model are defined
now everywhere except $U(\partial \Sigma)$. Let us consider such a
configuration, that for infinitely small contours $\cal C$ (winding around
$\Sigma$) the mentioned above Wilson loops are expressed as in (\ref{Sing}).
For $N \ne 0$ it is not possible to expand the definition of such a
configuration to $U(\partial \Sigma)$. However, this could become possible
within the unified model if the gauge group of the Standard Model $SU(3)\times
SU(2) \times U(1)/{\cal Z}$ is embedded into the simply connected group $\cal
H$. This follows immediately from the fact that any closed loop in such $\cal
H$ can be deformed smoothly to a point and this point could be moved to unity.
Actually, for such $\cal H$ we have $\pi_2({\cal H}/[SU(3)\times SU(2) \times
U(1)/{\cal Z}]) = \pi_1(SU(3)\times SU(2) \times U(1)/{\cal Z})$. This means
that in such unified model the monopole-like soliton states are allowed. The
configurations with (\ref{Sing}) and $N\ne 0$ represent fundamental monopoles
of the unified model\footnote{Actually these configurations were already
considered (see, for example \cite{Z6}, where they represent fundamental
monopoles of the $SU(5)$ unified model). However, in \cite{Z6} it was implied
that such soliton states could appear with the masses of the order of GUT scale
($10^{15}$ Gev). In our case the appearance of such objects is expected already
at the energies compared to $1$ Tev because we consider the unified model, in
which $\cal H$ is broken to the gauge group of the Standard Model already at
this scale.}. The other monopoles could be constructed of the fundamental
monopoles as of building blocks.  In the unified model, which breaks down to
the SM with the gauge group $SU(3)\times SU(2) \times U(1)$ such configurations
for $N\ne 0$ are simply not allowed. This gives us the way to distinguish
between the two versions of the Standard Model.

The unified model, which breaks down to the SM with the gauge group
$SU(3)\times SU(2) \times U(1)$ also contains monopoles because $\pi_2({\cal
H}/[SU(3)\times SU(2) \times U(1)]) = \pi_1(SU(3)\times SU(2) \times U(1)) =
Z$. They correspond to the Dirac strings with $ \int_{\cal C} B^{\mu} dx^{\mu}
= 6\pi K, K\in Z$ and should be distinguished from the monopoles (for
$N=1,2,3,4,5$) of the SM with the additional discrete symmetry via counting
their  magnetic fluxes.

Using an analogy with t'Hooft - Polyakov monopoles\cite{HooftPolyakov} we can
estimate masses of  the mentioned monopoles to be of the order of $ N
\sqrt{\frac{4\pi}{\alpha}}\Lambda \sim 40 N$ Tev, where $\Lambda \sim 1 \, {\rm
Tev}$ is the scale of the breakdown and $\alpha$
 is the fine structure constant ($\alpha(M_Z)\sim \frac{1}{128}$) (see, for example, \cite{Weinb},
 where monopoles were considered for an arbitrary compact simple gauge group in the BPS limit).
Here $N = 6$ for monopoles of the conventional Standard Model.

\section{Monopoles in Petite Unification models}

In order to illustrate the emergence of the additional $Z_3$ and $Z_2$
symmetries in the Standard Model we consider Petite Unification of strong and
Electroweak interactions discussed in \cite{PUT, PUT1}. In the mentioned papers
three possibilities to construct the unified theory at Tev were distinguished
among a number of various ones. Namely, let us consider the unified group to be
the product of $SU(4)_{PS}$ and $SU(N)^k$, where $SU(4)_{PS}$ unifies lepton
number with color as in Pati-Salam models \cite{PATI}. In the theory there are
two independent couplings $\alpha_s$ and $\alpha_W$ correspondent to the two
groups mentioned above. Then if we require that the spontaneous breakdown of
$SU(4)_{PS}\otimes SU(N)^k$ happens at a Tev scale we are left with the three
possibilities: ${\rm PUT}_0(N=2,k=4);\,{\rm PUT}_1 (N=2,k=3);\, {\rm
PUT}_2(N=3,k=2)$. The other choices of $N$ and $k$ cannot provide acceptable
values of coupling constants at the Electroweak scale. It should be mentioned
here that $\rm PUT_0$ seems to be excluded due to the extremely high value of
branching ratio for the process $K_L \rightarrow \mu e$.

It will be useful to represent the breakdown pattern correspondent to the
models $ PUT_1, PUT_2$ in terms of the loop variables $\Gamma, U$, and $\theta$
calculated along the arbitrary closed contour $\cal C$.

In $\rm PUT_2$ at the Electroweak scale $SU(4)_{PS}\otimes SU(3)^2$ parallel
transporter $\Omega$ along the contour $\cal C$ is expressed through $\Gamma,
U$, and $\theta$ as follows:

\begin{equation}
\Omega = \left( \begin{array}{c c}

\Gamma^+ e^{\frac{2i\theta}{3}} & 0  \\
0 & e^{-2i\theta}

\end{array}\right) \otimes\left( \begin{array}{c c c}

e^{\frac{-4i\theta}{3}} & 0 & 0 \\
0 & e^{\frac{2i\theta}{3}} & 0 \\
0 & 0 & e^{\frac{2i\theta}{3}}

\end{array}\right)\otimes\left( \begin{array}{c  c}

U e^{-\frac{i\theta}{3}} & 0  \\
0 & e^{\frac{2i\theta}{3}}

\end{array}\right)\label{PUT2}
\end{equation}

From (\ref{PUT2}) it is  straightforward that values (\ref{Sing}) of the Wilson
loops $\Gamma$, $U$, and $e^{i\theta}$ with $N = 0, 3 \in Z_2$ lead to $\Omega
= {\bf 1}$. The field strength of the $SU(4)_{PS}\otimes SU(3)^2$ gauge field
is expressed through $\Omega$ calculated along the infinitely small contour.
Then the pure gauge field action in the low energy limit (at the Electroweak
scale) is invariant under an additional $Z_2$ symmetry. This means that in $\rm
PUT_2$ actual breakdown pattern is $SU(4)_{PS}\otimes SU(3)^2 \rightarrow
SU(3)\times SU(2) \times U(1)/Z_2$ and not $SU(4)_{PS}\otimes SU(3)^2
\rightarrow SU(3)\times SU(2) \times U(1)$. Therefore, we expect $SU(2)/Z_2$
monopoles to exist in this unified model with the masses of the order of $120$
Tev.

Here we used the values of Electroweak charges calculated in \cite{PUT} in
order to represent the breakdown pattern in a form useful for our purposes. One
can check directly that the gauge group element of the form (\ref{PUT2}) acts
appropriately on the Standard Model fermions arranged in the representations
listed in \cite{PUT}. The same check could be performed also for the model $\rm
PUT_1$ considered below.

In $\rm PUT_1$ at the Electroweak scale $SU(4)_{PS}\otimes SU(2)^3$ parallel
transporter $\Omega$ along the contour $\cal C$ is expressed as follows:

\begin{equation}
\Omega = \left( \begin{array}{c c}

\Gamma^+ e^{\frac{2i\theta}{3}} & 0  \\
0 & e^{-2i\theta}

\end{array}\right)\otimes U \otimes \left( \begin{array}{c  c}

e^{-i\theta} & 0  \\
0 & e^{i\theta}

\end{array}\right) \otimes\left( \begin{array}{c  c}

e^{i\theta} & 0  \\
0 & e^{-i\theta}

\end{array}\right)\label{PUT1}
\end{equation}

It is  straightforward that values (\ref{Sing}) of the Wilson loops $\Gamma$,
$U$, and $e^{i\theta}$ with $N = 0, 2, 4 \in Z_3$ lead to $\Omega = {\bf 1}$.
This means that in $\rm PUT_1$ actual breakdown pattern is $SU(4)_{PS}\otimes
SU(2)^3 \rightarrow SU(3)\times SU(2) \times U(1)/Z_3$ and not
$SU(4)_{PS}\otimes SU(2)^3 \rightarrow SU(3)\times SU(2) \times U(1)$. Thus,
$SU(3)/Z_3$ monopoles should exist in this unified model with the masses of the
order of $80$ Tev.

\section{Lattice model for qualitative investigation of the monopoles}

 Basing on the
analogy with Nambu monopoles \cite{BVZ2006} we expect that the monopoles of the
models, which describe Tev - scale physics, may be condensed at $T > T_c$. In
order to investigate this phenomenon we construct the lattice toy model based
on
 $\rm PUT_2$.

We simplify $\rm PUT_2$ in such a way that the resulting model has the gauge
group $SU(3)$. At low energies parallel transporter $\Omega\in SU(3)$ along the
contour $\cal C$ is expressed through $U\in SU(2)$, and $\theta \in U(1)$ as
follows:

\begin{equation}
\Omega = \left( \begin{array}{c  c}

U e^{-\frac{i\theta}{3}} & 0  \\
0 & e^{\frac{2i\theta}{3}}

\end{array}\right)\label{PUT2l}
\end{equation}

In this theory the breakdown pattern is $ SU(3) \rightarrow  SU(2) \times
U(1)/Z_2$. Therefore,  $Z_2$ monopoles  exist. The lattice model contains link
$SU(3)$ field $\Omega$ and the adjoint scalar $\Phi \in su(3)$ defined on
sites. The potential for the scalar field may be considered in its simplest
form in the London limit. In this case the action has the form
\begin{eqnarray}
 S_g & = & \beta \!\! \sum_{\rm plaquettes}\!\!
 ((1-\mbox{${\small \frac{1}{3}}$} \, {\rm Re}\, {\rm Tr}\, \Omega_p )
 +\nonumber\\
 && + \gamma \sum_{xy}(1 - \frac{1}{2} \,{\rm Re}\, {\rm Tr}\, (\Omega_{xy} \lambda_8 \Omega^+ \lambda_8)),
\end{eqnarray}
where the plaquette variables are defined as $\Omega_p = \Omega_{xy}
\Omega_{yz} \Omega_{wz}^* \Omega_{xw}^*$ for the plaquette composed of the
vertices $x,y,z,w$. $\beta$ is usual $SU(3)$ coupling while $\gamma$ is
proportional to squared vacuum expectation value of the scalar field.

The hypercharge  $U(1)$ gauge field in the theory can be expressed through
$\Omega$ as follows: $\theta = \frac{3}{2}{\rm Arg}\, \Omega_{33}$. In lattice
theory monopole classical solution should be formed around the Dirac string,
 which is represented by the integer-valued
field defined on the dual lattice
\begin{equation}
 \Sigma = \frac{1}{2\pi}^*([d \phi]_{{\rm mod} 2\pi} - d \phi),
 \end{equation}
where $\phi = {\rm Arg}\, \Omega_{33}$. Here we used  notations of differential
forms on the lattice. For a definition of those notations see, for example,
~\cite{forms}. Then, worldlines of quantum $Z_2$ monopoles appear as the
boundary of the string worldsheet:
\begin{equation}
 j = \delta \Sigma
\label{jN}
\end{equation}

We expect, that the monopole currents should percolate above the transition
temperature $T_c$ in the finite temperature model. We expect that  these $Z_2$
monopoles resemble both $Z_3$ monopoles of realistic $\rm PUT_1$ and $Z_2$
monopoles of $\rm PUT_2$.

\section{Conclusions}

To conclude, we considered the Standard Model embedded into a unified model,
the symmetry of which breaks down to the gauge group of the SM at a few Tev.
During the breakdown monopoles may appear, which have masses of the order of
$40$ Tev. Those objects could become the lightest topologically stable magnetic
monopoles. In principle, they could be detected in future during high energy
collisions.

 We expect that in the early Universe at the
temperatures close to the temperature $T_c$ of the correspondent transition
 monopoles considered in this paper could appear in the elementary
processes with a high probability. Those monopoles may even be condensed at
$T>T_c$ as Nambu monopoles at the temperatures above the Electroweak transition
temperature \cite{BVZ2006}. We  present the lattice toy model for the
qualitative consideration of this phenomenon.

It is worth mentioning that the presence of monopoles considered here may be
excluded in the wide class of modern cosmological scenarios. In particular,
inflation theory implies reheating after inflation \cite{reheating}. Inflation
itself washes out relic magnetic monopoles created at the temperatures of the
order of GUT temperature $10^{15}$ Gev. In certain models the reheating
temperature is thought of to be between $10^6$ and $10^{10}$ Gev (see, for
example, \cite{leptogenesis_reheating} and references therein). After reheating
standard scenario of hot Universe works. The concentration of relic magnetic
monopoles in the Hot Big Bang scenario was estimated in \cite{Khlopov}.
According to this estimate the concentration of magnetic monopoles created at
$T \sim 1$ Tev appears to be essentially larger than the experimentally allowed
one.

I kindly acknowledge private communication with M.Yu. Khlopov. This work was
partly supported  by RFBR grants 06-02-16309, 05-02-16306, and 07-02-00237.


\begin{thebibliography}{99}



\bibitem{TEV}
J.A. Casas, J.R. Espinosa, I. Hidalgo, hep-ph/0607279

F. del Aguila, R. Pittau,  Acta Phys.Polon. B35 (2004) 2767-2780

\bibitem{BVZ2007}
B.L.G.~Bakker, A.I.~Veselov, and M.A.~Zubkov,  arXiv:0707.1017

\bibitem{little_higgs}
Martin Schmaltz, David Tucker-Smith, Ann.Rev.Nucl.Part.Sci. 55 (2005) 229-270

\bibitem{TopLittleHiggs}
Mark Trodden, Tanmay Vachaspati, Phys.Rev. D70 (2004) 065008


\bibitem{technicolor}
 Christopher T. Hill, Elizabeth H. Simmons,
 Phys.Rept. 381 (2003) 235-402;
Erratum-ibid. 390 (2004) 553-554

Kenneth Lane, hep-ph/0202255

R. Sekhar Chivukula, hep-ph/0011264

\bibitem{WS}
S.Weinberg, Phys.Rev.D 13, 974, 1976

L.Susskind, Phys.Rev.D 20, 2619, 1979

\bibitem{FS}
E.Farhi, L.Susskind, Phys.Rev.D 20, 3404, 1979

\bibitem{ExtendedTechnicolor}
 Thomas Appelquist, Neil Christensen, Maurizio Piai, Robert Shrock, Phys.Rev. D70 (2004) 093010

 Adam Martin, Kenneth Lane, Phys.Rev. D71 (2005) 015011

 Thomas Appelquist, Maurizio Piai, Robert Shrock, Phys.Rev. D69
(2004) 015002

Robert Shrock, hep-ph/0703050

 Adam Martin, Kenneth Lane, Phys.Rev. D71 (2005) 015011

\bibitem{PUT}
Andrzej J. Buras, P.Q. Hung, Phys.Rev. D68 (2003) 035015

\bibitem{PUT1}
Andrzej J. Buras, P.Q. Hung, Ngoc-Khanh Tran, Anton Poschenrieder, Elmar
Wyszomirski, Nucl.Phys. B699 (2004) 253

Mehrdad~Adibzadeh and P.Q.~Hung, hep-ph/0705.1154.

\bibitem{Z6}
C.Gardner, J.Harvey, Phys. Rev. Lett. {\bf 52} (1984) 879

Tanmay Vachaspati, Phys.Rev.Lett. 76 (1996) 188-191

Hong Liu, Tanmay Vachaspati, Phys.Rev. D56 (1997) 1300-1312

\bibitem{BVZ2003}
B.L.G.~Bakker, A.I.~Veselov, and M.A.~Zubkov, Phys. Lett. B {\bf  583}, 379
(2004);

\bibitem{BVZ2004}
B.L.G.~Bakker, A.I.~Veselov, and M.A.~Zubkov, Yad. Fiz. {\bf 68}, 1045 (2005).

\bibitem{BVZ2005}
B.L.G.~Bakker, A.I.~Veselov, and M.A.~Zubkov, Phys. Lett. B {\bf 620} (2005)
156-163.

\bibitem{BVZ2006}
B.L.G.~Bakker, A.I.~Veselov, and M.A.~Zubkov, Phys.Lett. B642 (2006) 147-152



\bibitem{Z6f}
K.S.~Babu, I.~Gogoladze, and K.~Wang, Phys. Lett. B {\bf 570}, 32 (2003);\\
K.S.~Babu, I.~Gogoladze, and K.~Wang, Nucl. Phys. B {\bf 660}, 322 (2003);







\bibitem{Z2007_2}
M.A.~Zubkov,   arXiv:0707.0731

\bibitem{Z2007}
M.A.~Zubkov, Phys. Lett. B {\bf  649}, 91 (2007).


\bibitem{HooftPolyakov}
Gerard 't Hooft, Nucl.Phys.B79:276-284,1974

Alexander M. Polyakov, JETP Lett.20:194-195,1974, Pisma
Zh.Eksp.Teor.Fiz.20:430-433,1974

\bibitem{Weinb}
Erick J. Weinberg, Nucl.Phys.B167:500,1980

\bibitem{PATI}
J.C.Pati, S.Radjpoot, A.Salam, Phys. Rev.  {\bf D 17}, 131 (1978)







\bibitem{forms}
M.I.~Polikarpov, U.J.~Wiese, and M.A.~Zubkov, Phys. Lett. B {\bf 309}, 133
(1993).


\bibitem{reheating}
 Lev Kofman , Andrei D. Linde  ,
Alexei A. Starobinsky, Phys.Rev.Lett.73:3195-3198,1994.

\bibitem{leptogenesis_reheating}
Rachel Jeannerot, Marieke Postma, JCAP 0512 (2005) 006


Masaaki Fujii, K. Hamaguchi, T. Yanagida, Phys.Rev. D63 (2001) 123513

M.Yu.Khlopov, A.D.Linde, Phys. Lett. (1984), V. 138B, PP. 265-268.

\bibitem{Khlopov}
Ya.B.Zeldovich, M.Yu.Khlopov, Phys.Lett. (1978), V. 79B, PP. 239-242



\end{thebibliography}
\end{document}